\documentclass[reprint,aps, prl, onecolumn,showpacs,groupedaddress,nofootinbib]{revtex4}  
\usepackage{graphics,graphicx}  
\usepackage{dcolumn}   
\usepackage{bm}        
\usepackage{amssymb}   
\usepackage{bigints}
\usepackage{xcolor}
\usepackage{tabularx}

\def\be{\begin{equation}}
\def\ee{\end{equation}}
\def\bea{\begin{eqnarray}}
\def\eea{\end{eqnarray}}

\def\br{{\mathcal B}}

\begin{document}
\title{ The $X(3872)$ tetraquarks in $B$ and $B_s$ decays. }
\author{Luciano Maiani, Antonio D. Polosa, Veronica Riquer}
\affiliation{Dipartimento di Fisica and INFN,  Sapienza  Universit\`a di Roma, Piazzale Aldo Moro 2, I-00185 Roma, Italy.}
\email{luciano.maiani@roma1.infn.it}
\email{antonio.polosa@roma1.infn.it}
\email{veronica.riquer@cern.ch}
\date{\today}
\begin{abstract}
We discuss how the latest data on $X(3872)$ in $B$ and $B_s $ decays speak about its tetraquark nature. 
The established decay pattern, including the up to date observations by CMS, are explained by the mixing of two quasi-degenerate, unresolvable, neutral states. 
The same mechanism also  explains isospin violations in $X$ decays and strongly suggests  that the lurking charged partners are required to have 
very small branching fractions in $J/\psi\, \rho^\pm$, well below the current experimental limits. In addition, a new prediction on the decay into $J/\psi\, \omega$ final states is attained.
The newest experimental observations are found to give thrust to the simplest tetraquark picture and  call for a definitive, in-depth study of final states with charged $\rho$ mesons.   
\end{abstract}

\pacs{14.40.Rt, 12.39.-x, 12.40.-y}

\maketitle
The discussion on the nature of $X(3872)$ has been going on, with conflicting conclusions, for about two decades since its first observation at Belle~\cite{Choi:2003ue}.

The $X(3872)$ is first of all a remarkable example of fine tuning realized in physics. Its mass is nearly equal to the sum of $D^0$ and $\bar D^{0*}$ open-charm mesons masses, whose composition of quantum numbers matches the $J^{PC}=1^{++}$ assigned to the $X(3872)$. This feature is not met at the same level by any one of the so called `exotic' resonances discovered over the years. Reviews on exotic hadrons can be found in~\cite{Chen:2016qju,Esposito:2016noz,Ali:2017jda,Guo:2017jvc,Lebed:2016hpi,Olsen:2017bmm,book}. 

Despite its decay modes involving the $J/\Psi$, $X(3872)$ cannot be interpreted as a pure charmonium state. One of the simplest reasons for this is due to the fact that it decays in $J/\psi \, \rho$ and $J/\psi \, \omega$ with similar rates, thus violating isospin.

The proximity of the $X(3872)$ to the $D^0\bar D^{0*}$ threshold, isospin violations, and the lack of evidence so far of a complete multiplet of charged and neutral states, has convinced a large part of the community working on this problem that the $X(3872)$ should be a sort of deuteron made of neutral $D$ mesons, namely a  $D^0\bar D^{0*}$ molecule, with a very small binding energy, which is still unknown because of the uncertainties in the determination of the $X(3872)$  mass value. 
On the other hand the $X(3872)$ is produced, with a very large cross section, at proton-(anti)proton colliders in regions of  transverse momenta of final state hadrons, which are  too high (above $p_T\sim$ 15~GeV) for the formation of such a loosely-bound molecule~\cite{Bignamini:2009sk,Esposito:2015fsa}, see also~\cite{Brodsky:2015wza}.

Alternatively one might suppose that only color forces determine the
structure of the $X(3872)$, which is often referred as to the compact
tetraquark interpretation. Loosely bound molecules and compact
tetraquarks are  the two opposite extrema of a spectrum of more complex
solutions that the problem may have.  This not to mention that some
authors consider the possibility that the $X$ might simply be a threshold
kinematical effect, a cusp, as detailed in \cite{Swanson:2006st}.
Another interesting suggestion, pursued by Voloshin and collaborators
(see for example \cite{Li:2013ssa}), is that of hadrocharmonia, {\it i.e.} 
relatively  compact charmonia embedded in a light quark mesonic excitation.

The compact tetraquark model was developed in~\cite{Maiani:2004vq,Maiani:2014aja,Maiani:2017kyi}. It proposes that  $X(3872)$ belongs to a complex of four-quark bound states: $X_u,~X_d$ and $X^\pm=[cu][\bar c\bar d],[cd][\bar c\bar u]$ where parentheses mark diquark correlations.

Such states are expected to be very close in mass to each other.  In a first estimate, Ref.~\cite{Maiani:2004vq} gave a $X_d-X_u$ separation close to $2(m_d-m_u)\sim 7$~MeV. However, a second state close to $X(3872)$ has not been observed, and  upper bounds have been given for the branching ratios of $B$ meson decays into $X^\pm$~\cite{Aubert:2004zr,Choi:2011fc}.   
Building on the analysis of isospin breaking hadron masses~\cite{Rosner:1998zc,Karliner:2017gml}, which takes into account the effect of the electromagnetic interactions, it was suggested~\cite{Maiani:2017kyi} that  $X_u$ and $X_d$
are much closer in mass  than  expected, so as to be two unresolved lines inside the $J/\psi\,  \pi^+\pi^-$ peak. This quasi-degeneracy is reached assuming a separation of scales between the diquark size and the size of the whole diquark-antidiquark composite state, also considered in~\cite{Esposito:2018cwh}. To the best of our knowledge, the possibility of a diquark-antidiquark repulsion was first mentioned by Selem and Wilczek in \cite{Selem:2006nd}.
Another result obtained  in~\cite{Maiani:2017kyi} was  that $X_u-X_d$ mixing, estimated from the branching ratios of $X(3872)\to J/\psi + 2\pi$ or $3\pi$, would push the branching ratio for the production of $X^\pm$ in $B$ meson decays well below the experimental limits of~\cite{Aubert:2004zr,Choi:2011fc}, thus calling for more refined searches.

\begin{figure}[htbp]
   \centering
   \includegraphics[width=0.3 \linewidth]{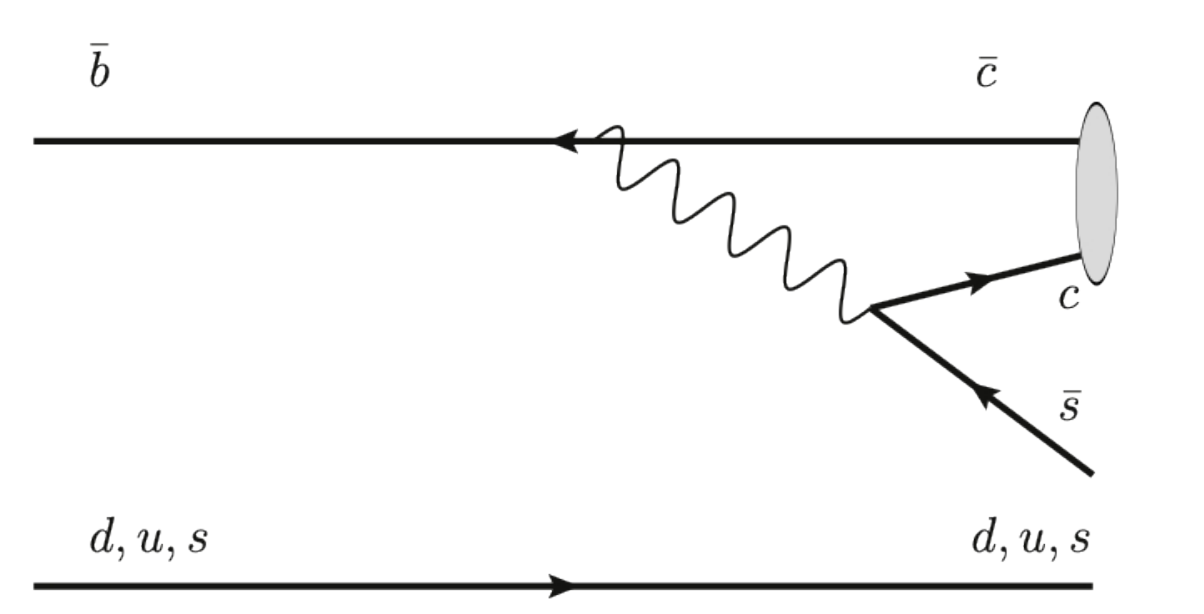}
   \caption{The valence quarks in $B$ and $B_s$ decays. A pair of sea quarks is formed in the blob to generate the $X$ tetraquarks.}
         \label{decay}
\end{figure}

CMS has recently reported~\cite{Sirunyan:2020qir} a determination of the branching ratio of the weak decay
\be
\br (B^0_s\to \phi X(3872)  \to \phi\, J/\psi\, \pi^+ \pi^-)=(4.14 \pm 0.54~({\rm stat.}) \pm 0.32~({\rm syst.}) \pm 0.46~(\br)) \times 10^{-6}\label{bs}
\ee
Comparing to other similar decays, the following pattern is observed~\cite{Sirunyan:2020qir,Tanabashi:2018oca} 
\be
\br ( B^0_s\to \phi\, X \to \phi \, J/\psi\, \pi^+ \pi^-)\simeq \br ( B^0\to K^0 X  \to K^0\, J/\psi\, \pi^+ \pi^-)\simeq
\frac{1}{2}\br ( B^+\to K^+\, X  \to K^+\, J/\psi\, \pi^+ \pi^-)\label{pattern}
\ee
We will show how this pattern clearly emerges from the simplest decay diagram in Fig.~\ref{decay} in the compact tetraquark picture of the $X(3872)$. 
In addition, the pattern in~\eqref{bs} and~\eqref{pattern}, combined with our previous analysis~\cite{Maiani:2017kyi} of the branching fractions of $X(3872)\to J/\psi +2\pi/3\pi$, allows to determine uniquely mixing and couplings of the two tetraquarks $X_u=[cu][\bar c\bar u],~X_d=[cd][\bar c\bar d]$. From these results we derive two new predictions
\begin{enumerate}
\item The branching ratio of the decays of $B$ mesons into $J/\psi + 3\pi $
 \be
R^{+0}_{3\pi}=\frac{\br ( B^+\to K^+ X(3872)  \to K^+ J/\psi ~\pi^+ \pi^-\pi^0)}{\br ( B^0\to K^0 X(3872)  \to K^0 J/\psi ~\pi^+ \pi^-\pi^0)}=0.87\pm0.06\label{res1}
\ee
\item A definite range for the production of the charged tetraquark $X^\pm$ in $B$ decays~\footnote{In the loosely bound molecular model, X(3872) has no charged partners, see {\it e.g.} Ref.~\cite{Guo:2017jvc}.}

 \bea
 &&0.05<R^-_{2\pi}=\frac{\br (B^0\to K^+ X(3872)^- \to K^+ J/\psi~\pi^0\pi^-)}{\br (B^0\to K^0 X(3872) \to K^0 J/\psi~\pi^+\pi^-)}<0.57\label{res2}
 \eea
 to be compared with the present limit $R^-_{2\pi}<1$~\cite{pdg}. 
  
  \end{enumerate}
  These predictions can be tested experimentally and, if supported, would provide a decisive clarification on the nature of the $X(3872)$.

\begin{figure}[htbp]
   \centering
   \includegraphics[width=0.7 \linewidth]{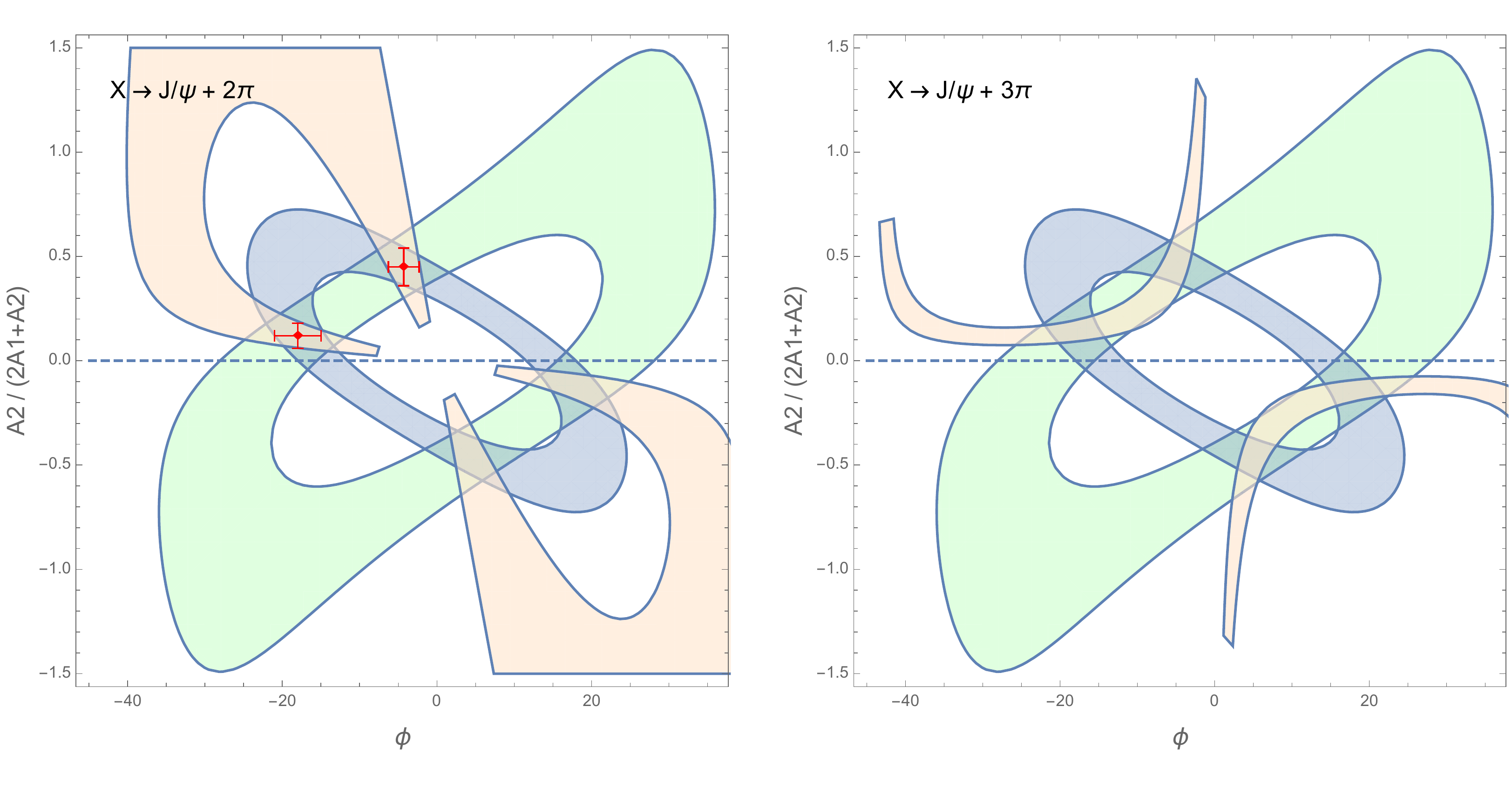}
   \caption{Left-panel: Intersecting regions in the $\phi-z$ plane corresponding to the observed $R^{+0}_{2\pi}$, $R(B^0)$ and $R(B^+)$ ratios. Right-panel: Same as for left-panel, for  $R^{+0}_{3\pi}$, $R(B^0)$ and $R(B^+)$ ratios.}
         \label{mixing}
\end{figure}

Assuming a tetraquark $X(3872)$, in the blob of Fig.~\ref{decay} one has to create a light quark pair from the sea. The overall weak decay is
\be
\Big({\bar b} + u,d,s\Big)_{B^+,B^0,B_s} \longrightarrow {\bar c} + c {\bar s}+(d {\bar d} ~{\rm or}~u{\bar u})_{{\rm sea}}+u,d,s \nonumber
\ee
The decays $B^{0,+}\to X~K^{0,+}$ are then described by two amplitudes:  $A_1$, where the $\bar s$ forms the Kaon with the spectator $u$ or $d$ quark, and $A_2$, where it forms the Kaon with a $d$ or $u$ quark from the sea. In terms of the unmixed states
\bea
&& {\cal A}(B^0\to X_d\, K^0)\sim A_1+A_2\notag\\
\label{unox}
&& {\cal A}(B^0\to X_u\, K^0)\sim  A_1\\
&& {\cal A}(B^0\to X^-\,K^+) \sim A_2\notag
\eea 
and
\bea
&&{\cal A}(B^+\to X_d\, K^+)\sim A_1\notag\\
\label{duex}
&& {\cal A}(B^+\to X_u\, K^+)\sim A_1+A_2\\
&& {\cal A}(B^+\to X^+\,K^0 ) \sim A_2\notag
\eea 
With near degeneracy of $X_{u,d}$, even a small $q\bar q$ annihilation amplitude inside the tetraquark could produce sizeable mixing. We consider the mass eigenstates in the isospin basis, namely
\bea
&&X_1=\cos \phi~ \frac{X_u+X_d}{\sqrt{2}}+\sin \phi~\frac{X_u-X_d}{\sqrt{2}}\nonumber \\
&&X_2=-\sin \phi~ \frac{X_u+X_d}{\sqrt{2}}+\cos \phi~\frac{X_u-X_d}{\sqrt{2}}\label{mixmass}
\eea
(we can take $\cos\phi>0$, so that $-\pi/4<\phi<+\pi/4$). It is straightforward\footnote{In Eqs.~(18) and (19) of Ref.~\cite{Maiani:2017kyi}, one should correct the typo: $p_\rho/p_\omega\to p_\omega/p_\rho $.} to compute the rate for $B$ going to $X(3872)$, the sum of two unresolved, almost degenerate lines, followed by decay into $J/\psi+2\pi/3\pi$, as function of the mixing angle $\phi$ and of the ratio of the isospin zero and isospin one amplitudes, $2A_1+A_2$, $A_2$, respectively. 
The result~\cite{Maiani:2017kyi} is reported in the two panels of Fig.~\ref{mixing} by the donut-shaped regions, which correspond to the experimental values of the two ratios~\cite{pdg}
\bea
&&R(B^0)=\frac{\Gamma (B^0\to K^0~X(3872)\to K^0\, J/\psi\, 3\pi)}{\Gamma (B^0\to K^0~X(3872)\to K^0\, J/\psi\,2\pi)}=1.4\pm 0.6 \\
&&R(B^+)=\frac{\Gamma (B^+\to K^+\, X(3872)\to K^+ \,J/\psi\,3\pi)}{\Gamma (B^+\to K^+ ~X(3872)\to K^+\,J/\psi\, 2\pi)}=0.7\pm 0.4
\eea
Let us now turn to  the results~\eqref{bs} and \eqref{pattern}. From Eqs.~\eqref{unox} to \eqref{mixmass}, and recalling that 
\be
{\cal A}(X_{1,2} \to J/\psi\, \rho) = \sin \phi,\cos \phi
\ee
one easily finds the ratio of the $B^+$ to $B^0$ rates  in~\eqref{pattern}. The result is
\bea
&&R^{+0}_{2\pi}=\frac{\br ( B^+\to K^+ \,X(3872)  \to K^+\, J/\psi \,\pi^+ \pi^-)}{\br ( B^0\to K^0\, X(3872)  \to K^+\, J/\psi \,\pi^+ \pi^-)}=\frac{1+3z^2-(1-z^2)\cos(4\phi)-2z\sin(4\phi)}{1+3z^2-(1-z^2)\cos(4\phi)+2z\sin(4\phi)}\\
&&\text{where}\quad z=\frac{A_2}{2A_1+A_2}\label{rat+02pi}\notag
\eea
A few observations are in order
\begin{enumerate}
\item We have summed over the rates of $X_1$ and $X_2$, as required by the hypothesis~\cite{Maiani:2017kyi} that the two neutral states are both within the $J/\psi \rho$ width.
\item  $R^{+0}_{2\pi}=1$ if either $\phi$ or $z$ vanish, see \eqref{unox} and \eqref{duex}.
\item The periodicity in $\phi$ of \eqref{rat+02pi} is $\pi/2$, coinciding with the range of physically different configurations in  \eqref{mixmass}.
\item $2A_1+A_2$ and $A_2$ correspond to isospin $0,1$ and their relative sign is inessential; we may take $z>0$ by convention.
\end{enumerate}
Using the experimental branching ratios~\cite{pdg} and adding errors in quadrature, we find
\be
R^{+0}_{2\pi}=2.0\pm0.6\label{ratpz}
\ee
The corresponding region in $\phi,z$ space is reported in Fig.~\ref{mixing} left-panel.  The value~\eqref{ratpz} is in remarkable agreement with the previous determination based on the $2\pi$ vs. $3\pi$ decays. It leads to the two solutions marked with points and bars in Fig~2, left-panel\footnote{The solutions with $z<0$ are simply reflections of $z>0$ ones and do not correspond to physically different solutions.}
\bea
&&{\rm {Solution~1}}:\phi=-18^\circ \pm3^\circ\quad ~z=0.12\pm0.06\notag\\
&&{\rm Solution~2}:\phi=-4.3^\circ \pm 2^\circ \quad z=0.45\pm0.09 \label{solutions}
\eea

For $B^0_s$ decay, only the spectator quark can lead to the $\phi$ meson in the final state. The decay is described by one amplitude,  $A_3$, with the same role as $A_1$ in $B^0$ decay 
\be
R^{s0}_{2\pi}=\frac{\br ( B_s\to \phi \, X(3872)  \to \phi \, J/\psi \, \pi^+ \pi^-)}{\br ( B^0\to K^0\, X(3872)  \to K^+\, J/\psi\, \pi^+ \pi^-)}=\left(\frac{A_3}{A_1+A_2/2}\right)^2 \frac{2\sin(2\phi)^2}{1+3z^2-(1-z^2)\cos(4\phi)+2z\sin(4\phi)}\label{ratbs}
\ee
From~\eqref{solutions} and \eqref{ratbs} we find
\bea
&&R^{s0}_{2\pi}({\rm Solution~1})=\left(\frac{A_3/A_1}{1.14}\right)^2 \times 1.35=1~{\rm for}~A_3/A_1=0.97;\notag\\
&&R^{s0}_{2\pi}({\rm Solution~1})=\left(\frac{A_3/A_1}{1.82}\right)^2 \times 0.08=1~{\rm for}~A_3/A_1=6.5
\eea
On the other hand, the near  equality of the branching ratios~\cite{pdg}
\bea
\br ( B^0\to K^{*0}\, X(3872)\to K^{*0}\, J/\psi\, \pi^+\pi^-)&=&(4.0\pm 1.5)\times 10^{-6}\notag\\
\br ( B^0\to K^{0}\, X(3872)\to K^{0}\, J/\psi\, \pi^+\pi^-)&=&(4.3\pm 1.3) \times10^{-6}
\eea
suggests that the couplings $A_{3,4}$ for  $B$ decay into vector mesons, are similar to $A_{1,2}$ of~\eqref{unox} and~\eqref{duex}. Barring unforeseen cancellations, we may conclude that the CMS pattern \eqref{pattern} selects Solution 1 over Solution 2
\footnote{A quantitative conclusion can be found by determining  $A_{3,4}$ from the decays $B\to K^* X(3872)\to K^* J/\psi~ \pi^+\pi^-$ as done here for decays into K mesons. The branching ratio of the decay  $B^+\to K^{*+} X(3872)$ is not available at present~\cite{pdg}.}.
This fact has a simple interpretation. In Solution 1, $z$ is small and the mixing is such that the contribution of $X_u$ dominates in $B^0$ decay. Thus, to a good approximation, meson formation in $B^0$ decay is dominated by the spectator quark as in $B_s$ decay. 


Using the parameters of Solution 1, one obtains the two predictions in~\eqref{res1} and~\eqref{res2}. We conclude that  the new results by CMS mark an advancement in the understanding of the $X(3872)$ problem and call for a few more steps to do on the experimental side, to safely decide among existing interpretations.


\end{document}